# Pauli Master Equation numerical analysis of coherent and incoherent dressed fermions in triplet unconventional superconductors

Pedro L. Contreras E.*

Departamento de Física, Universidad de Los Andes, Mérida, 5101, Venezuela.

ORCID: https://orcid.org/0000-0002-3394-1195

E-mail: pcontreras@ula.ve

*Corresponding author

**Abstract:** We report two types of dressed fermions in a triplet superconductor with an in-situ disorder effective field. They are obtained numerically by analyzing with the Pauli Master Equation, the self-consistent imaginary part data of the elastic-scattering cross-section. We use a two-component irreducible representation of the order parameter with quasi-point nodes and study the quasi-classical effective probabilistic density distribution as function of disorder. We find a stable coherent quantum state of dressed fermions, and slow characteristic decay time for dilute disorder. Also, we find an incoherent quantum state with diffused dressed fermions and enriched disorder with self-consistency increasing decoherence, a fast characteristic decay time, and a diffuse unitary resonance. We conclude that the most stable dressed fermion states are those in which the field has dilute disorder, the threshold zero gap, and some dressed fermions behave like an s-wave superconductor, showing a tiny gap.

**Keywords**: Quasi-classical effective probabilistic density distribution. Effective disorder field. Coherent dressed fermions. Diffused dressed fermions. Quasi-point nodes order parameter. Unconventional triplet superconductors.

**Introduction**

In this work, we apply the results of the imaginary part of the elastic scattering cross-section to the analysis of the quasi-classical probabilistic density distribution $\mathcal{W}(t)$ in the context of the Master Pauli equation [1], and calculate the time decay for fermionic dressed wave functions. All these within the context of a two component triplet time reversal breaking (TRSB) order parameter (OP) in an unconventional superconductor, that belongs to the irreducible representation $E_u(\Gamma_5^-)$, as a function of (in-situ) stoichiometric disorder.

Additionally, we propose to use for averages of fermions quantities, the distribution function $f(t)$ used in the collision integral $I$ and the $\tau$-approximation of the Boltzmann kinetic equation applied to different solid state phenomena in classical textbooks [2-8], but using the imaginary part of the elastic scattering cross-section instead of a constant inverse collision lifetime as derived in [8].

In this simulation, we input two different values of the zero superconducting gap ($\Delta_0$) to contrast the results for the cuasi classical probability density $\mathcal{W}(t)$, and its behavior in the presence of stoichiometric (in-situ) disorder. We use numerical data obtained from simulating two zero superconducting gaps [9, 10] with tight-binding bonds and a quasi-point nodal structure (QP). We also use the fundamental kinetic relation $1/\tau[\widetilde{\omega}(\omega)] = \nu[\widetilde{\omega}(\omega)] = 2\,\Im[\widetilde{\omega}(\omega)]$ [11], That shows a direct relation among the imaginary part of the elastic scattering cross-section ($\Im$), the inverse lifetime ($\tau^{-1}$), and the collision frequency ($\nu$) of fermion quasiparticles dressed by a effective disordered self-consistent field.

This cuasi-classical problem arises when we formulate the question about the role of disorder in the coherence or incoherence of dressed fermions wave functions in unconventional superconductors. We use a differential equation for the probability density function as formulate in [1]. The equation for the quasi-stationary probability density function $\mathcal{W}(t)$ belongs to non-equilibrium statistical mechanics, and it was originally formulated by Prof. W. Pauli in 1928. For our work, it is given by the equation

$$\left(\partial\,\mathcal{W}(t)/\partial\,t\right)_{qsd} + 1/\tau\,\mathcal{W}(t) = 0. \qquad (1)$$

We substitute the inverse scattering lifetime of its imaginary elastic cross-section in equation 1, and we get the effective Pauli Master Equation [11]

$$\left(\partial\,\mathcal{W}_{\widetilde{\omega}}(t)/\partial\,t\right)_{qsd} + 2\,\Im[\widetilde{\omega}(\omega)]\mathcal{W}_{\widetilde{\omega}}(t) = 0, \qquad (2)$$

where $\mathcal{W}_{\widetilde{\omega}}(t)$ is the quasi-stationary effective probabilistic density distribution. If the imaginary effective self-consistent term vanishes at higher frequencies, then a steady-state kinetic equation 3 emerges [12]. The effective self-consistent equation 1 is difficult to solve numerically, and we need to prove the results with different physical examples, as the one we are going to solve in this manuscript.

$$\left(\partial\,\mathcal{W}_{\widetilde{\omega}}(t)/\partial\,t\right)_{qsd} = 0. \qquad (3)$$

On the other hand, the threshold decay time of the quasi-stationary probability density function is given by the number

$$\mathcal{W}_{th}(t^*) = 1/e \approx 0.3679. \tag{4}$$

The characteristic decay time *t\** is given by the expression

$$\mathcal{W}(t^*) = |\psi_{\widetilde{\omega}}(t)|^2 = 1/e\ \mathcal{W}(0) = 1/e\ |\psi_{\widetilde{\omega}}(0)|^2 = \mathcal{W}_{th}(t^*)\ \mathcal{W}(0) = \mathcal{W}_{th}(t^*) \tag{5}$$

$\mathcal{W}(t^*)$ tell us the time in what the initial probability density function changes its initial shape by $1/e$. This is due to the interaction of a particular dressed fermion wave-function with others fermions, and the effective self-consistent field. An explanation for the characteristic decay-time values is the following:

- If the constructed physical model has a slow decay, it has a long characteristic time *t\** (from 3.0 meV$^{-1}$ – 4.0 meV$^{-1}$), a quantum coherent state.
- If the physical model decays fast, it has a short characteristic decay time *t\** (less than 1.0 meV$^{-1}$), a quantum decoherent state.

On the other hand, we should notice that although we do not use the Keldish formalism [2, 13] for non-equilibrium properties, it is possible to use the Abrikosov-Gorkov formalism that includes normal dirty-dressed zero-temperature Green functions for the fermions elastic cross-section [14-16] not so far for the thermodynamical equilibrium [2], and use these results to calculate the self-consistent probability density depending on disorder.

We start the derivation of the solution for equation 1 empirically, by writing in rationalized Planck units the time depending quantum mechanical self-consistent wave-function $\psi_{\widetilde{\omega}}(t) \sim e^{-(iE_{\widetilde{\omega}} + \Im[\widetilde{\omega}(\omega)])\,t}$, where we introduce the imaginary part of the elastic scattering cross-section as the self-consistent quasiclassical damping $\Im[\widetilde{\omega}(\omega)]$. In non-relativistic quantum mechanics, the quasi-classical damping has the meaning of jumping from a quantum state to another [1].

Thus, we have a dressed-fermion wave-function, and its complex conjugate given by the expressions

$$\psi_{\widetilde{\omega}}(t) \sim e^{-(iE_{\widetilde{\omega}} + \Im[\widetilde{\omega}(\omega)])\,t}\ \&\ \psi^*_{\widetilde{\omega}}(\widetilde{\omega}, t) \sim e^{+(iE_{\widetilde{\omega}} - \Im[\widetilde{\omega}(\omega)])\,t} \tag{6}$$

If $t > 0$, the solution of equation 1 for the probability density distribution is a function of the imaginary part of $\widetilde{\omega}$. We write it as a function of stoichiometric (in-situ) disorder ($\Gamma^+$) given by equation 7

$$\mathcal{W}_{\widetilde{\omega}}(t) = \psi_{\widetilde{\omega}}(t)\,\psi^*_{\widetilde{\omega}}(\widetilde{\omega}, t) = |\psi_{\widetilde{\omega}}(t)|^2 = \mathcal{W}_{\widetilde{\omega}}(0)\,e^{-2\,\Im[\widetilde{\omega}(\omega)]t} \tag{7}$$

We suppose that the self-consistent density probability at zero characteristic time is given by the value $\mathcal{W}_{\widetilde{\omega}}(0) = 1$. Henceforth, we obtain a differentiable equation 7 that exponentially decays at different rates of stoichiometric (in-situ) disorder.

Other case, that we not solve numerically in this manuscript, but it is important to briefly review is to use the same approach for $\mathcal{W}_{\widetilde{\omega}}(t)$, but using the Boltzmann kinetic equation in the τ-approximation [2, 5] using a self-consistent scattering elastic lifetime. We write the well-known Boltzmann equation

$$df/dt = (\partial f(t)/\partial t)_{coll} + \vec{v}\, \partial f/\partial \vec{r} + e/m\, \vec{E}\, \partial f/\partial \vec{p} = I. \qquad (8)$$

The right term $I$ is the collision integral, $f(t)$ is the distribution function. In the left side we have $\vec{v}$ & $e/m\, \vec{E}$, the electrons velocity & an external electric field, respectively.

In the case of fermion quasiparticles, equation 8 has been solved analytically in different situations of physical interest [2-8]. One classical example is the anomalous skin effect [17], and its analytical solution using equation 8. The solution can be obtained with singularities in a 3D Fermi surface [18,19]. For $\omega\, \tau\, (\omega) \geq 1$ the Fermi surface singularities are very sensitive to Fermi liquid interactions [19]. But, in the unitary scattering limit for a time-breaking triplet superconductor, the effective self-consistent field becomes very sensitive to disorder, therefore it holds that $\tilde{\omega}\, \tau\, (\tilde{\omega}(\omega)) \geq 1$, and $\ell a^{-1} \sim 1$.

If we neglect the velocity dependence ($\vec{v}\, \partial f/\partial \vec{r}$), and the electrical external field ($e/m\, \vec{E}\, \partial f/\partial \vec{p}$) in equation 8, we are left with a simplified time-dependent Boltzmann equation. Additionally, we use the τ-approximation for the collision integral $I$, which is given by the expression

$$I = -1/\tau\, (f - f_0). \qquad (9)$$

In order to study effective self-consistent fields, we change the constant $1/\tau$ term arising in normal state metals, for the imaginary [5] damping as function of an effective field, i.e., $1/\tau_s = 2\, \gamma\, [\tilde{\omega}(\omega)]$.

Therefore, we obtain a kinetic equation in the presence of an effective self-consistent field [11]

$$df/dt = (\partial f(t)/\partial t)_{coll} + 2\, \gamma\, [\tilde{\omega}(\omega)](f) = 0. \qquad (10)$$

Equation 10 implies dressed fermion collisions, with three different solutions depending on the sign of the time, but we use one solution, equation 11,

$$f(t) = f_0 + (f(o) - f_0) e^{2\, \gamma\, [\tilde{\omega}(\omega)]} \qquad (11)$$

We ought to notice that the classical mechanical damping $\gamma[\tilde{\omega}(\omega)]$ is the negative of the imaginary elastic cross-section $\Im\, [\tilde{\omega}(\omega)]$. Thus,

$$f(t) = f_0 + (f(o) - f_0) e^{-2\, \Im\, [\tilde{\omega}(\omega)]} \qquad (12)$$

where the letter $f_0$ is the distribution function at equilibrium, and $f(o)$ is the distribution function at $t = 0$ in equation 12.

In addition, for equations 7 and 12, we can use a TB parametrization [20] to introduce five microscopic parameters in the first neighbors approximation inside the function $\Im\, [\tilde{\omega}(\omega)]$: The Fermi energy and the hopping constant (linked to the Fermi average), the collision parameter $c$ (associated with the strength of the effective self-consistent field), the disorder linked to the strontium lattice vibrations, when the linear momentum is transferred from the dressed fermions to the Sr atoms, and finally the zero superconducting gap [10].

We remark that for this analysis, the imaginary term shows the separation among frequency-dependent unconventional superconductors. It means, TRSB triplet superconductors with the irreducible representation $E_u(\Gamma_5^-)$ with a real frequency window of $\pm 4$ meV for the analysis of the cross-section [10, 11, 21, 22], or HTSC superconductors with strontium impurities, a real frequency window of $\pm 150$ meV, and the irreducible representation $B_{1g}$ [23].

This work is organized as follows: First part is a brief introduction to the Pauli Master equation and the Boltzmann kinetic equation, the second section is a parameter framework for the unconventional model of a superconductor, third part uses the effective field data to analyze the Pauli Master equation, and the fifth part states the conclusions.

**Self-Consistent Effective Framework**

For the numerical analysis of the effective field from the elastic scattering cross-section, we use a model of the crystal strontium ruthenate considering for this study to have a triplet pairing [24-30]. In addition, we use an elastic cross-section effective field formalism firstly proposed to study heavy fermions [31-33]. It is very important to recall that a strong disorder dependence on strontium ruthenate was proposed for the first time in [34].

We use a TB mode for the superconductor strontium ruthenate with a Fermi level $\epsilon_F = -0.4$ meV, which leaves a QP gap around the $(0, \pm \pi)$ and $(\pm \pi, 0)$ Brillouin points [9, 10]. However, if the Fermi energy is $\epsilon_F = -0.04$ meV, there are point nodes, and the analysis of this work does not hold [21]. The effective self-consistent fermions Green-functions treatment of the imaginary part of the elastic scattering cross-section is performed using rationalized Planck units ($\hbar = k_B = 1$). Therefore, the real and imaginary axes are given in meV units and the characteristic decay time in meV$^{-1}$ units. The readers are encouraged to check how to observe unitary effective cross-section similarities in fermions [22], and ultra-cold Bose gases [35]. The model for the first neighbor TB normal fermions dispersion is given according to

$$\xi(k_x, k_y) = \epsilon_F + 2\,t\,[\cos(k_x\,a) + \cos(k_y\,a)]. \tag{13}$$

Henceforth, in this work the nodal configuration and the Fermi average are controlled by the first neighbors Fermi energy with the value $\epsilon_F = -0.4$ meV, and a first neighbor hoping parameter $t = 0.4$ meV. In this case, the mean free path $l$ (a key non-equilibrium statistical mechanical parameter) is in the unitary limit when $l\,k_F \sim l\,a^{-1} \sim 1$, where $k_F$ is the Fermi momentum vector and $a$ is the square lattice parameter. Additionally, we use the 2D irrep. $E_u(\Gamma_5^-)$ with first neighbors. The triplet TRSB OP in this case is a $z$-vector that shows two order parameter structures: QP and point nodes [21,36]

$$\mathbf{\Delta}(k_x, k_y) = \Delta_0[(\sin(k_x a) + i\sin(k_y a)]\hat{z}. \tag{14}$$

The experimental threshold zero superconducting gap $\Delta_0$ is 1.0 meV. Several experimental works, and fits found that strontium ruthenate has a threshold superconducting zero value where holds that $\Delta_0 \leq 1.0$ meV [37-42]. The resonant unitary limit needs c = 0, and an imaginary elastic scattering cross-section with the form [8]

$$\Im [\widetilde{\omega}(\omega + i\, 0^+)] = \pi\, \Gamma^+ \frac{1}{g(\widetilde{\omega})}. \tag{15}$$

The disorder parameter of the effective field is $\Gamma^+ = n_{stoich}/(\pi^2 N_F)$ and the scattering strength is $c = 1/(\pi N_F U_0)$ [43]. The imaginary term is proportional to the first power of the disorder parameter $\Gamma^+$ as is deeply analyzed in [44]. In a series of works, some previously mentioned, we have numerically studied the effective self-consistent field for the superconducting phase in strontium ruthenate using a set of five parameters for the quasi-point and point nodes OP. All these studies were performed in the unitary and intermedium scattering regimes, because there is no numerical evidence of hydrodynamic scattering in the strontium ruthenate crystal [10,11,21,22].

In the following section, to construct the effective quantum mechanical wave functions using self-consistent data, we perform simulations with the following values. The scattering parameter taking the following values $c = 0$. The stoichiometric disorder varies from a dilute disorder $\Gamma^+ = 0.05$ meV to a very enriched disorder $\Gamma^+ = 0.35$ meV. Additionally, we use two values for the zero superconducting gap $\Delta_0$. The threshold experimental value for strontium ruthenate ($\Delta_0 = 1.00$ meV), and an intermedium gap ($\Delta_0 = 0.40$ meV).

**Numerical Results**

Figure 1 shows the first simulation of the imaginary part of the scattering cross-section for the unitary case (equation 15 that implies $c = 0$), with a quasi-point nodal structure (QP), and where the isolated nodes are located at $(0, \pm\pi)$ and $(\pm\pi, 0)$ Brillouin points with a TRSB irrep. $E_u(\Gamma_5^-)$. In Figure 1, we additionally use the superconducting zero threshold experimental gap $\Delta_0 = 1.0$ meV, the Fermi energy parameter $\epsilon_F = -0.4$ meV, and the hopping constant $t = 0.4$ meV. We add a third axis on the left of Figure 1 to plot the five values of the in-situ disorder $\Gamma^+$, that is, from a dilute disorder $\Gamma^+ = 0.05$ meV (blue plot), to a very enriched disorder $\Gamma^+ = 0.35$ meV (violet plot).

The colors of the plots and disorder values are the following: blue-dilute disorder, orange-quasi-optimal disorder, green-optimal disorder, red-quasienriched disorder, violet-very enriched disorder. In Figure 1 it is seen that from the dilute $\Gamma^+ = 0.05$ meV simulation (colored blue) to an quasienriched disorder with $\Gamma^+ = 0.20$ meV (colored red), there is a similar behavior in the shape of the three figures, i.e., a central well-defined unitary resonance followed by a continuous curve with one minimum, which ends with the normal state line.

The tiny gap [44] that resembles a s-wave superconductor [45] happens for $\Gamma^+ = 0.05$ meV, and it is seen within the blue line. The very enriched disorder plot with $\Gamma^+ = 0.35$ meV is colored violet, and it does not show a remarkable minimum shape, but shows an spread (diffuse) unitary resonance, and a normal state line. These two different behaviors will be analyzed with the help of the Pauli Master equation below.

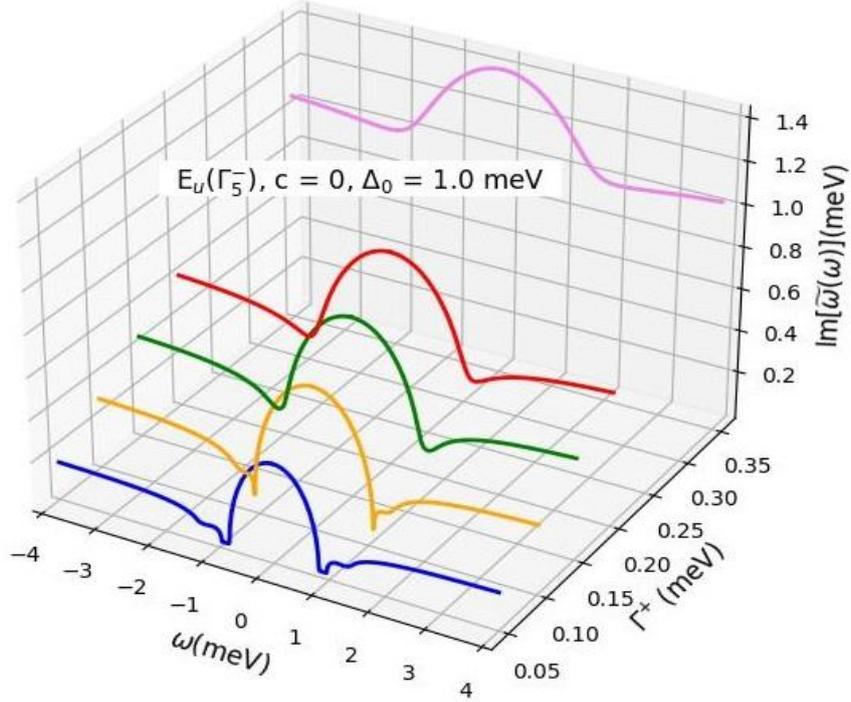

*Figure 1. Imaginary part of the elastic scattering cross-section for a threshold experimental zero gap in the unitary limit, and five values of in-situ disorder for a TRBS with an irrep. $E_u$ ($\Gamma^-_5$) OP.*

Figure 2 shows the result for the solution of equation 2 using equation 7. The exponential decay of the quasi-stationary effective probability density function for dressed fermions $\mathcal{W}_{\tilde{\omega}}(t)$ as a function of the time $t$ has a decreasing behavior as disorder increases, and $\Delta_0 =$ 1.0 meV. To perform this simulation, we used the set of data shown in Figure 1. The colors for each plot in Figure 2 correspond to the same colors for the disorder values in Figure 1.

We clearly see that a smallest negative slope for $\mathcal{W}_{\tilde{\omega}}(t)$ as a function of $t$ happens for the blue plot that represents the exponential decay of the tiny gap plot, the tiny gap is clearly seen below a 1.0 meV in Figure 1 with $\Gamma^+ = 0.05$ meV. We find the biggest negative slope for $\mathcal{W}_{\tilde{\omega}}(t)$ as function of $t$ (almost an asymptotic behavior) when we input a very enriched disorder (violet colored plot with $\Gamma^+ = 0.35$ meV) and the shape is a spread unitary resonance.

Figure 2 shows an interesting feature. The linear behavior of the plot $\mathcal{W}_{\tilde{\omega}}(t)$ as a function of time $t$ is seen to be different for the in-situ disorder values. Less disorder means more linear time with negative slope in Figure 2. As disorder increases, $\mathcal{W}_{\tilde{\omega}}(t)$ shows a non-linear behavior for smaller times $t$. Summarizing, we say that the linear time behavior decreases, as the disorder rate $\Gamma^+$ increases for the effective probabilistic density function $\mathcal{W}_{\tilde{\omega}}(t)$. Additionally, the tiny gap case (blue plot) has more linear time behavior, and therefore seems to be the most coherent quantum state. It occurs for dressed fermions with dilute disorder. Finally, in Figure 2 when the characteristic time $t > 4.0$ meV$^{-1}$, the quasi-stationary effective probabilistic function $\mathcal{W}_{\tilde{\omega}}(t)$ tends to be a constant.

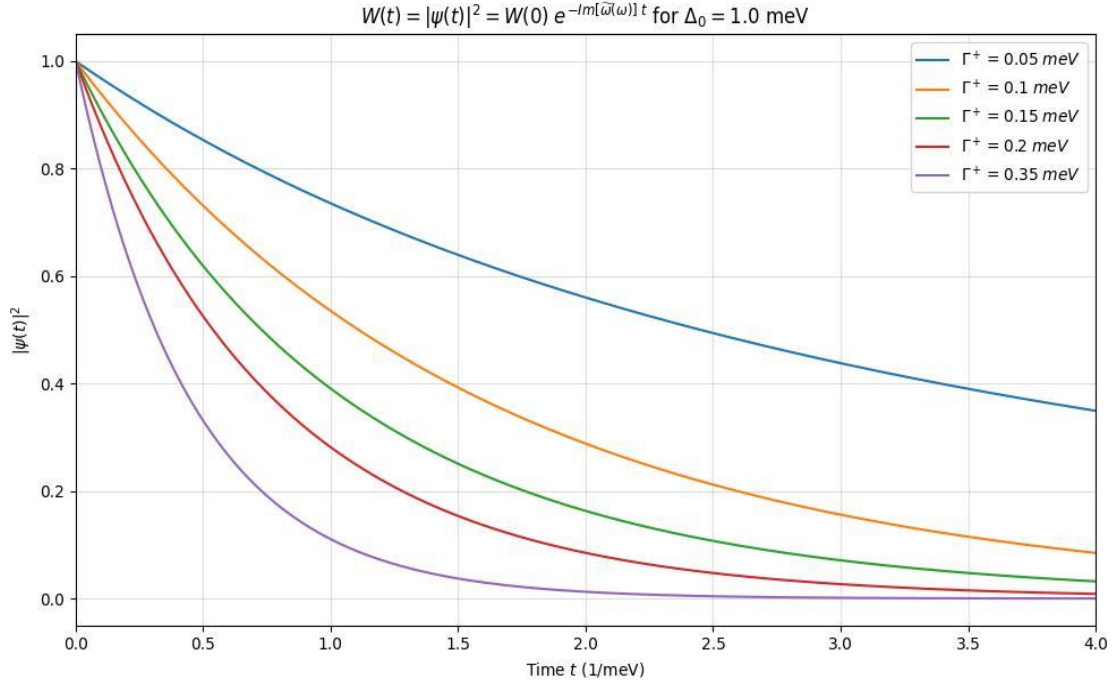

*Figure 2. An effective quasi-stationary probability density as a function of time (in rationalized Planck units) numerically found with the data from Figure 1, using several amounts of in-situ disorder & the threshold gap.*

Figure 3 shows the second simulation of the imaginary part of the scattering cross-section for the unitary case (equation 15 & $c = 0$), with the same quasi-point nodal OP, but in Figure 3, we use the superconducting zero gap $\Delta_0 = 0.4$ meV, and the same parameters for the Fermi energy, and the hopping constant.

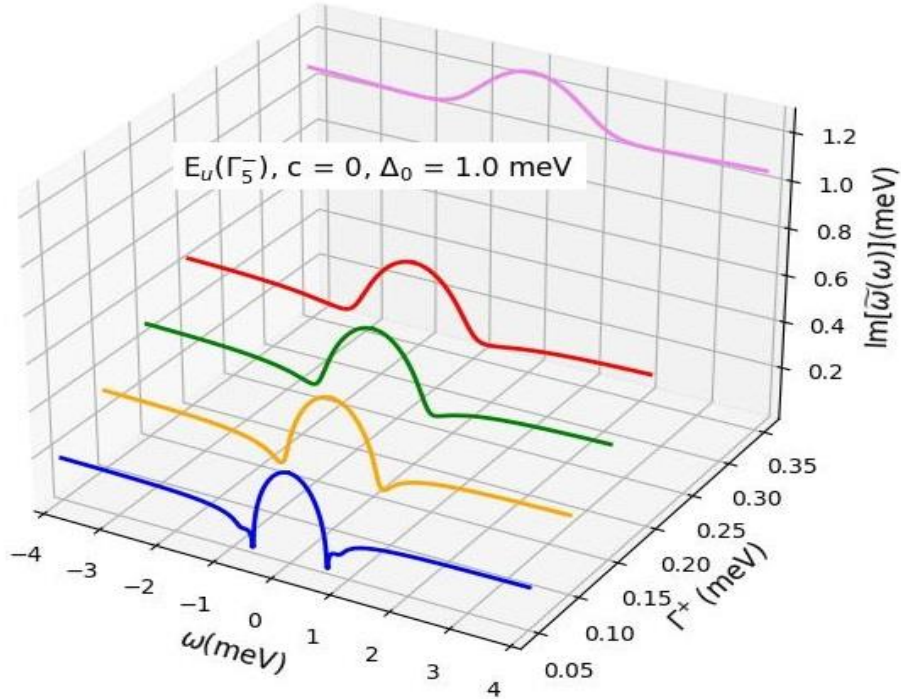

*Figure 3. Imaginary parts of the elastic scattering cross-section for a zero 0.4 meV gap, and five values of disorder for an TRBS irrep. $E_u$ ($\Gamma^-_5$) OP.*

In Figure 3, the third axis on the left of the figure also shows the amount of disorder. The plots in Figure 3 with five values of disorder $\Gamma^+$ have the same colors as in Figure 1 & Figure 2. Disorder varies from a dilute-$\Gamma^+ = 0.05$ meV (blue plot) to a very enriched-$\Gamma^+ = 0.35$ meV (violet plot). The very enriched disorder plot in Figure 3 (colored violet) it does not show a minimum, but shows a diffuse unitary resonance, and a normal state line. This behavior will be analyzed below.

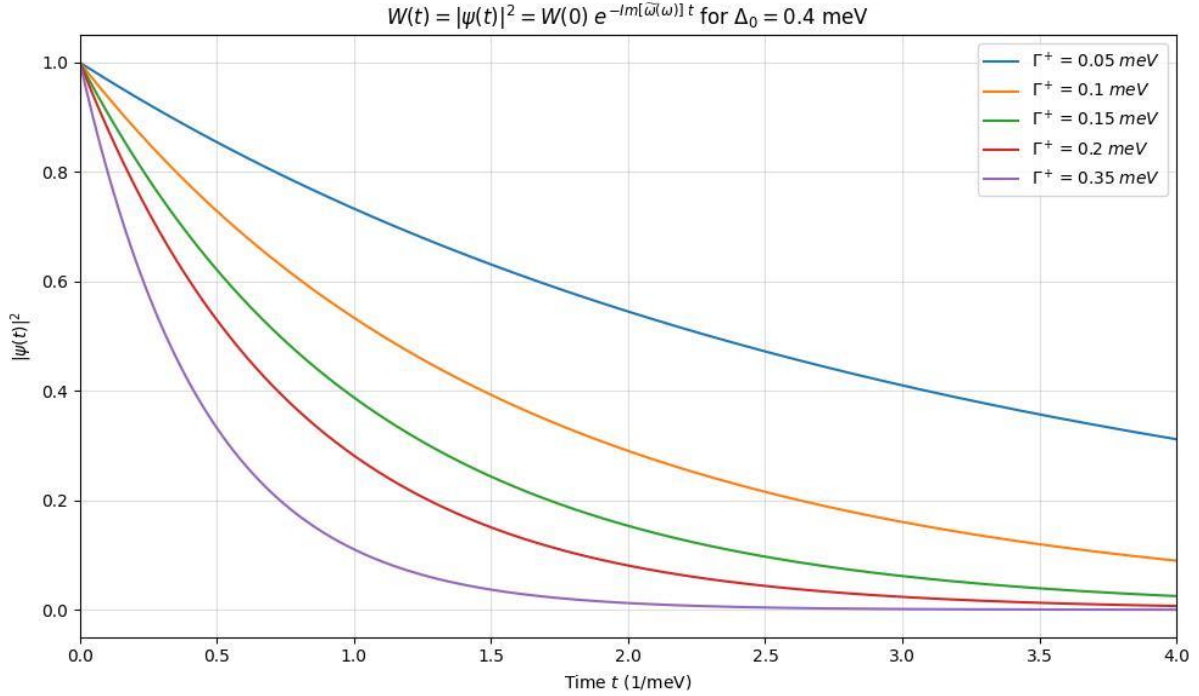

*Figure 4. An effective quasi-stationary probability density as a function of time (in rationalized Planck units) numerically found with the data from Figure 3, using several amounts of disorder and a 0.4 meV superconducting zero gap.*

Figure 4 shows the numerical results for the quasi-stationary probability density function $\mathcal{W}_{\widetilde{\omega}}(t)$ as a function of time $t$ given by equation 7, when we input a zero superconducting gap $\Delta_0 = 0.4$ meV. To perform this simulation, we use the set of data shown in Figure 3. The colors of each plot in Figure 4 correspond to the same effective disorder values with their respective colors in Figures 1-3.

In Figure 4, we see an almost identical $\mathcal{W}_{\widetilde{\omega}}(t)$ behavior as we see in Figure 2, i.e., a smallest negative slope for the effective quasi-stationary probability density $\mathcal{W}_{\widetilde{\omega}}(t)$ as a function of small values of time, happens for the blue plot, which represents the results given by the imaginary elastic cross-section with a minimum point and a dilute in-situ disorder colored blue in Figure 3 (this minimum is clearly seen below the real frequency of 1.0 meV). The biggest negative slope for the effective quasi-stationary probability density $\mathcal{W}_{\widetilde{\omega}}(t)$ as function of small times, occurs for the very enriched disorder as in Figure 2, it is also colored violet, and it has an almost asymptotic behavior. In Figure 4 happens the same behavior as in Figure 2, that is, for t > 4.0 meV$^{-1}$, $\mathcal{W}_{\widetilde{\omega}}(t)$ becomes a constant.

Henceforth, the dressed fermions effective probability density $\mathcal{W}_{\widetilde{\omega}}(t)$ analysis given by Figure 4 shows a similar result to the analysis given in Figure 2. This is not a coincidence, since $\Delta_0 = 0.4$ meV is the best fit for the experimental electronic thermal conductivity data [46] for the γ-sheet of the strontium ruthenate Fermi surface [47].

Next step, we calculate and plot the characteristic decay time $t^*$ as a function of in-situ disorder $\Gamma^+$ to find the difference between the two $\mathcal{W}_{\widetilde{\omega}}(t)$ plots presented in Figures 2 & 4.

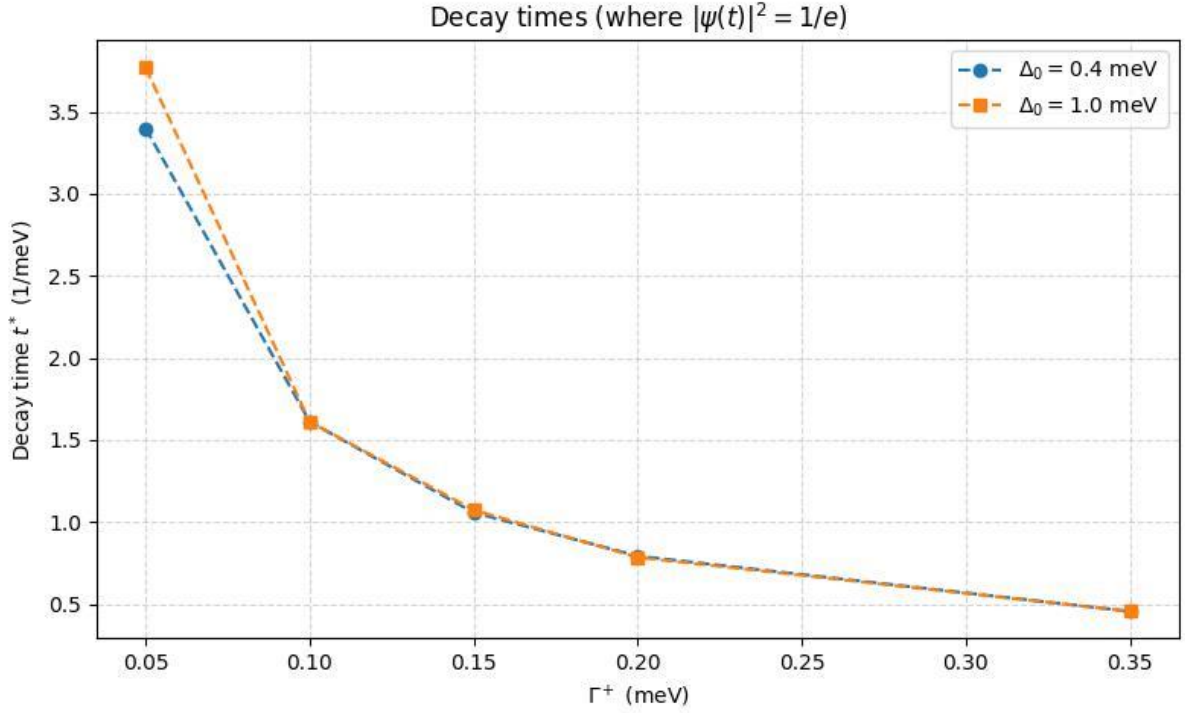

*Figure 5: Characteristic decay time as a function of disorder for the two zero superconducting gaps.*

Figure 5 sketches the characteristic decay time for both zero supercoducting gaps. The square orange plot represents the data for the zero threshold superconducting gap $\Delta_0 = 1.0$ meV, and the circular blue plot shows the calculation for the zero superconducting gap value $\Delta_0 = 0.4$ meV. The 0Y axis represents the characteristic decay time $t^*$, and the axis 0X shows the in-situ effective-$\Gamma^+$ disorder. We recall that $t^*$ is the time in which the system will lose its initial quantum coherence by certain constant amount and it is given by equation 4. Therefore, in Figure 5 both plots (orange & blue) show the same tendency, i. e., $t^*$ decreases as the amount of disorder increases.

| Disorder value $\Gamma^+$ (meV) | Characteristic decay time $t^*$ (meV$^{-1}$) for $\Delta_0 = 1.0$ meV | Characteristic decay time $t^*$ (meV$^{-1}$) for $\Delta_0 = 0.4$ meV |
|---|---|---|
| 0.05 | 3.767535 | 3.398788 |
| 0.10 | 1.611222 | 1.611222 |
| 0.15 | 1.074148 | 1.058116 |
| 0.20 | 0.785571 | 0.793587 |
| 0.35 | 0.456914 | 0.456914 |

*Table 1: Characteristic decay time for different values of stoichiometric disorder, and zero superconducting gaps for a TRBS irrep. $E_u$ ($\Gamma^-_5$) OP.*

Summarizing, Figure 5 and Table 1 show that:

- If there are big values in the 0Y axis, the dressed fermions have a slow decay time; thus, they keep its initial quantum state for more time and they have a well-defined coherent quantum state. In other words, the dressed fermions avoid the decoherence brought by their own interaction and the self-consistent effective field. The shape of the unitary resonance is well-defined.

- If the 0Y axis has a small value, the dressed fermions decay fast, and have a quantum decoherent state, because they interact strongly with other dressed fermions, and the disorder brought by the effective self-consistent field. The unitary resonance shows a diffusive behavior.
- Both zero superconducting gaps have the same plot tendency (to decrease as function of the characteristic time). However, the orange plot for the s-wave like gap has the highest characteristic time (3.767535 meV$^{-1}$) show in Table 1. Thus, it is the most coherent quantum state.
- Both cases show that for a very enriched amount of disorder, they are decoherent quantum states, more diffuse dressed fermions, and a they have a characteristic time (0.456914 meV$^{-1}$) show in Table 1.

**Conclusions**

This work analyzes the quantum coherent and quantum decoherent behavior [1,48,49] of dressed fermions in the case of having a TRBS unconventional superconducting state with a irrep. $E_u(\Gamma_5^-)$ order parameter for two different numerical zero gap values using the Pauli Master equation as it is formulated in [1]. The simulation is performed with the input of a self-consistent effective field data obtained from the imaginary part of the elastic scattering cross-section. We calculate the effective quasi-classical fermionic probabilistic density distribution $\mathcal{W}_{\tilde{\omega}}(t)$, and the characteristic decay time $t^*$ for different disorder values.

Calculations are performed in the unitary regime, where the imaginary part of the elastic cross-section has a resonance. In the many body theory of strontium ruthenate, the self-consistent effective field appears when the quantum state of dressed fermions and the disordered strontium field, determine each other. Strontium distort the fermions cloud, generating the effective internal field and modifying the energy levels.

We identify two cases:

- First, the self-consistent field reinforces a quantum stable coherent state, which reduces the coupling to the field. Thus, dressed fermions become isolated, keeping a long quantum coherence, and slow characteristic time decay in the frequency regions where the self-consistency field decreases. Thus, the first class of fermions are robust coherent fermions with a characteristic decay time of 3.767535 meV$^{-1}$, some of them localized inside the tiny s-superconducting gap, they are realized if the disorder is dilute.
- Second, if the self-consistent effective field induces interactions with dressed fermions, their coupling and decoherence increase. They have a rapid characteristic decay time. Therefore, the second class of fermions are incoherent fermions with a characteristic decay time of 0.456914 meV$^{-1}$, localized in a spread unitary resonance. We call them, diffuse-dressed fermions, and they exist with enriched disorder.

Concluding, we state that the vector [50] 2D TRSB irrep. $E_u(\Gamma_5^-)$ OP [9, 25] simulates some dressed fermion properties of the strontium ruthenate crystal, and it has valuable information about its physical behavior, because the superconducting state is frequency-sensitive to some parameters that can be analyze using a self-consistent effective field, with the help of the imaginary part of the elastic scattering cross-section, and the Pauli Master Equation (PME).

Thus, discovered 31 years ago [51,52], strontium ruthenate is still a fascinating topic of experimental, theoretical, & numerical research, that partially support the two-components model, i.e., all physical kinetic low temperature experiments, shear stress, ultrasound, thermodynamic, pressure, and elastocaloric physical effects, respectively [30, 53-59].

Respect to the $B_{1g}$ irrep. [60] of the OP for strontium ruthenate, we think that is not an adequate irrep. Although it has a single component, line nodes, and reduces the temperature as nonmagnetic disorder increases, in terms of the elastic scattering cross-section, it describes well another HTSC compound, the strontium-doped lanthanum cuprate [61] using the imaginary part of the self-consistent effective field [23 ,62, 63].

With this manuscript, we emphasize the importance of joint non-relativistic quantum mechanics & non-equilibrium statistical mechanics to study unconventional superconductors.

**Data Availability Statement**

The data that support the findings of this study are available on request from the corresponding author. The data are not publicly available due to privacy or ethical restrictions

**Authorship Declaration Statement**

Pedro Contreras: Conceptualization, Methodology, Software, Investigation, Visualization, Validation, Writing – original draft.

**Funding**

No funding was received for this manuscript.

**Declaration of Competing Interest**

The author declares that he has no known competing financial interests or personal relationships that could have appeared to influence the work reported in this paper.

**References**

[1] Kvashnikov, I. 2003, The Theory of Systems out of Equilibrium, Moscow State University Press.
[2] Pitaevskii, L., Lifshitz, E. & Sykes, J. 1981, Physical Kinetics, Vol. 10. Pergamon Press, ISBN 0-08-020641-7
[3] Reif, F. 1965, Fundamentals of Statistical and Thermal Physics, McGraw Hill, ISBN 0-07-051800-9
[4] Daily, J. 2019, Statistical Thermodynamics, an Engineering Approach, Cambridge University Press, https://doi.org/10.1017/9781108233194
[5] Blatt, F. 1957, Theory of Mobility of Electrons in Solids, Academic Press.
[6] Ashcroft, N. & Mermin, N. 1976, Solid State Physics, Holt, Rinehart and Winston, ISBN 978-0-030-83993-1
[7] Ziman, J. 1979, Models of Disorder: The Theoretical Physics of Homogeneously Disordered Systems, Cambridge University Press, ISBN-10-0521292808
[8] Mineev, V. & Samokhin, K. 1999 Introduction to Unconventional Superconductivity, Gordon and Breach Science Publisher ISBN: 90-5699-209-0
[9] Miyake, K. & Narikiyo, O. Model for unconventional superconductivity of Sr2RuO4, effect of impurity scattering on time-reversal breaking triplet pairing with a tiny gap. 1999. Phys. Rev. Lett. 83:1423, https://doi.org/10.1103/PhysRevLett.83.1423

[10] Contreras, P., Osorio, D. & Ramazanov, S. Nonmagnetic tight-binding effects on the γ-sheet of Sr2RuO2. 2022. Rev. Mex. Fís. 68(2):020502, https://doi.org/10.31349/RevMexFis.68.020502

[11] Contreras, P. A constant self-consistent scattering lifetime in superconducting strontium ruthenate. 2024. Rev. Mex. Fís., vol. 70(6) pp. 060501 1-9 DOI: https://doi.org/10.31349/RevMexFis.70.060501

[12] Jansen, R., Farid, B. & Kelly, M. The steady-state self-consistent solution to the nonlinear Wigner-function equation; a new approach. 1991. Physica B: Condensed Matter, Vol. 175(1–3):49, https://doi.org/10.1016/0921-4526(91)90688-B

[13] Keldysh, L. Diagram technique for non-equilibrium processes. 1965. JETP 20(4):1018. ISSN: 1090-6509

[14] Abrikosov, A. 1972. Introduction to the Theory of Normal Metals, Academic Press, ISBN 13: 9780126077728

[15] Abrikosov, A. & Gorkov, L. Contribution to the Theory of Superconducting Alloys with Paramagnetic Impurities. 1961. JETP 12, 1243, ISSN 1090-6509

[16] Abrikosov, A., Gorkov L. & Dzyaloshinski I. 2012. Methods of Quantum Field Theory in Statistical Physic. Dover Books on Physics, ISBN-13 978-0-486-63228-5

[17] Reuter, G. & Sondheimer, E. The theory of the anomalous skin effect in metals. 1948. Proceedings of the Royal Society A. 195:336-364 https://doi.org/10.1098/rspa.1948.0123

[18] Kaganov, MI., Lyubarskiy, G. & Mitina, A. The theory and history of the anomalous skin effect in normal metals. 1997. Physics Reports. 288(1-6):291-304. https://doi.org/10.1016/S0370-1573(97)00029-X

[19] Kaganov, MI. & Contreras, P. Theory of the anomalous skin effect in metals with complicated Fermi surfaces.1994. JETP 106(6):985, ISSN: 1090-6509

[20] Harrison, W. 1989. Electronic Structure and Properties of Solids, Dover Books on Physics, ISBN: 9780486660219

[21] Contreras, P., Osorio, D. & Tsuchiya, S. Quasi-point versus point nodes in Sr2RuO2, the case of a flat tight binding. 2022. γ-sheet. Rev. Mex. Fis. 68(6):060501 https://doi.org/10.31349/RevMexFis.68.060501

[22] Contreras, P., Osorio, D. & Devi, A. The effect of nonmagnetic disorder in the superconducting energy gap of strontium ruthenate. 2022. Physica B: Condensed Matter. 646:414330, https://doi.org/10.1016/j.physb.2022.414330

[23] Contreras, P. Osorio, D. & Devi A. Self-Consistent Study of the Superconducting Gap in the Strontium-doped Lanthanum Cuprate, International Journal of Applied Mathematics and Theoretical Physics. 2023. Vol. 9(1), pp. 1-13, https://doi.org/10.11648/j.ijamtp.20230901.11

[24] Maeno, Y., Hashimoto, H., Yoshida, K., Nishizaki, S., Fujita, T., Bednorz, J. & Lichtenberg, F. Superconductivity in a layered perovskite without copper. 1994. Nature, 372:532-534. https://doi.org/10.1038/372532a0

[25] Rice, T. & Sigrist, M. Sr2RuO4: an electronic analogue of $^3$He? 1995, Journal of Physics, Condensed Matter. 7(47): L643-L648.

[26] Luke, G., Fudamoto, Y., Kojima, K., Larkin, M., Merrin, J., Nachumi, B., Uemura, Y., Maeno, Y., Mao, Z., Mori, Y., Nakamura, H. & Sigrist, M. Time-reversal symmetry-breaking superconductivity in Sr2RuO4. 1998. Nature, 394(6693):558-561 https://doi.org/10.1038/29038

[27] Ishida, K., Mukuda, H., Kitaoka, Y., Asayama, K., Mao, ZQ., Mori, Y. & Maeno, Y. Spin-triplet superconductivity in Sr2RuO4 identified by $^{17}$O Knight shift. 1998, Nature, 396:658-660, https://doi.org/10.1038/25315

[28] Duffy, J., Hayden, S., Maeno, Y., Mao, Z., Kulda, J. & McIntyre, G. Polarized-neutron scattering study of the Cooper-pair moment in Sr2RuO4. 2000. Phys, Rev. Lett. 85(25):5412-5415, https://doi.org/10.1103/PhysRevLett.85.5412

[29] Bergemann, C., Mackenzie, A., Julian, S. J., Forsythe D. & Ohmichi E. Quasi-two-dimensional Fermi liquid properties of the unconventional superconductor Sr2RuO4. 2003. Advances in Physics, 52:7, 639-725, 10.1080/00018730310001621737

[30] Mackenzie, A. & Maeno, Y. The superconductivity of Sr2RuO4 and the physics of spin-triplet pairing. 2003. Reviews of Modern Physics. 75(2):657-712, https://doi.org/10.1103/RevModPhys.75.657

[31] Pethick, C. & Pines, D. Transport processes in heavy-fermion superconductors. 1986. Phys Rev Lett. 57(1):118-121, https://doi.org/10.1103/PhysRevLett.57.118

[32] Hirschfeld, P., Wölfle, P. & Einzel, D. Consequences of resonant impurity scattering in anisotropic superconductors: Thermal and spin relaxation properties. 1988. Phys. Rev. B. 37(1):83, https://doi.org/10.1103/PhysRevB.37.8

[33] Balatsky, A. Salkola, M. & Rosengren. A. 1995, Impurity-induced virtual bound states in d-wave superconductors. 1995. Phys. Rev. B 51(21):15547-15551, https://doi.org/10.1103/physrevb.51.15547

[34] Mackenzie, A., Haselwimmer, R., Tyler, A., Lonzarich, G., Mori, Y., Nishizaki, S. & Maeno, Y. Extreme dependence of superconductivity on the disorder in Sr2RuO4. 1998. Phys. Rev. Lett. 80, 161, https://doi.org/10.1103/PhysRevLett.80.161

[35] Pitaevskii, L. Superfluid Fermi liquid in a unitary regime. 2008. Phys. Usp. 51, 603, https://doi.org/10.3367/UFNr.0178.200806i.0633

[36] Walker, MB. & Contreras, P. Theory of elastic properties of Sr2RuO4 at the superconducting transition temperature. 2002. Phys. Rev. B. 66(21):214508, https://doi.org/10.1103/PhysRevB.66.214508

[37] Lupien, C., MacFarlane, W., Proust, C., Taillefer, L., Mao, Z. & Maeno, Y. Ultrasound attenuation in Sr2RuO4: An angle-resolved study of the superconducting gap function. 2001. Phys. Rev. Lett. 86(26):5986-5989, https://doi.org/10.1103/PhysRevLett.86.5986

[38] Contreras, P., Walker, MB. & Samokhin, K. Determining the superconducting gap structure in Sr2RuO4 from sound attenuation studies below Tc. 2004. Phys. Rev. B. 70(18):184528, https://doi.org/10.1103/PhysRevB.70.184528

[39] Suzuki M., Tanatar, M., Kikugawa, N., Mao, Z., Maeno, Y. & Ishiguro, T. Universal heat transport in Sr2RuO4. 2002. Phys. Rev. Lett., 88:227004, https://doi.org/10.1103/PhysRevLett.88.227004

[40] Agterberg, D. Rice, T. & Sigrist, M. Orbital dependent superconductivity in Sr2RuO4. 1997. Phys. Rev. Lett. 78:3374. https://doi.org/10.1103/PhysRevLett.78.3374

[41] Zhitomirsky, M. & Rice, T. Interband proximity effect and nodes of superconducting gap in Sr2RuO4. 2001. Phys. Rev. Lett. 87(5):057001. https://doi.org/10.1103/PhysRevLett.87.057001

[42] Deguchi, K., Mao, Z., Yaguchi, H. & Maeno, Y. Gap structure of the spin-triplet superconductor Sr2RuO4 determined from the field-orientation dependence of the specific heat. 2004. Phys. Rev. Lett., 92:047002. https://doi.org/10.1103/PhysRevLett.92.047002

[43] Schachinger, E. & Carbotte J. Residual absorption at zero temperature in d-wave superconductors. 2003. Phys. Rev. B 67, 134509 https://doi.org/10.1103/PhysRevB.67.134509

[44] Lifshitz, I, Gredeskul, S. & L. Pastur, A. 1988. Introduction to the theory of disordered systems John Wiley and Sons, ISBN-10: 0471875333

[45] Cooper, L. Bound electron pairs in a degenerate Fermi gas. 1956. Phys. Rev. 104 (4), 1189–1190. https://doi.org/10.1103/PhysRev.104.1189

[46] Tanatar, M., Nagai, S., Mao, Z., Maeno, Y., & Ishiguro, T. Thermal conductivity of superconducting Sr2RuO4 in oriented magnetic fields. 2001. Phys. Rev. B 63: 064505 https://doi.org/10.1103/PhysRevB.63.064505

[47] Contreras, P. Electronic heat transport for a multiband superconducting gap in Sr2RuO4. 2011. Rev. Mex. Fis. 57 (5) pp. 395-399. ISSN: 2683-2224

[48] Pauli, W. 1928. Festschrift on the occasion of his 60th birthday A. Sommerfeld. Hirzel, Leipzig, 30.
[49] Novakovic, B. & Knezevic, I. 2011. Quantum Master Equations in Electronic Transport. in: Vasileska, D. & Goodnick, S. (eds) Nano-Electronic Devices. Springer, New York, NY. https://doi.org/10.1007/978-1-4419-8840-9_4
[50] Larkin, A. Vector pairing in superconductors of small dimensions. 1965. JETP Letters. 2(5):105 ISSN: 0370-274X
[51] Maeno, Y., Ikeda, A. & Mattoni, G. Thirty years of puzzling superconductivity in $Sr_2RuO_4$. 2024. Nat. Phys. 20, 1712–1718, https://doi.org/10.1038/s41567-024-02656-0
[52] Leggett, A. & Liu, Y. Symmetry Properties of Superconducting Order Parameter in $Sr_2RuO_4$. 2021. J. Sup. Nov. Magn. 34:1647, https://doi.org/10.1007/s10948-020-05717-6
[53] Hicks, C., Brodsky, D., Yelland, E., Gibbs, A., Bruin, J., Barber, M., Edkins, S., Nishimura, K., Yonezawa, S., Maeno, Y. & Mackenzie, A. Strong increase of Tc of $Sr_2RuO_4$ under both tensile and compressive strain. 2014. Science. 344(6181):283-285, https://doi.org/10.1126/science.1248292
[54] Benhabib, S., Lupien, C., Paul, I., Berges, L., Dion, M., Nardone, M., Zitouni, A., Mao, Z., Maeno, Y., Georges, A., Taillefer, L. & Proust, C. Ultrasound evidence for a two-component superconducting order parameter in $Sr_2RuO_4$. 2020. Nature Physics, https://doi.org/10.1038/s41567-020-1033-3
[55] Ghosh, S., Shekhter, A., Jerzembeck, F., Kikugawa, N., Sokolov, D., Brando, M., Mackenzie, A., Hicks, C. & Ramshaw, B. Thermodynamic evidence for a two-component superconducting order parameter in $Sr_2RuO_4$. 2020. Nature Physics, https://doi.org/10.1038/s41567-020-1032-4
[56] Grinenko, V., Das, D., Gupta, R. et al. Unsplit superconducting and time reversal symmetry breaking transitions in $Sr_2RuO_4$ under hydrostatic pressure and disorder. 2021. Nat. Commun. 12, 3920, https://doi.org/10.1038/s41467-021-24176-8
[57] A. Steppke, A. Pustogow, et al. Upper critical field of $Sr_2RuO_4$ under in-plane uniaxial pressure. 2024. Phys. Rev. B 107 064509, https://doi.org/10.1103/PhysRevB.107.064509
[58] Jerzembeck, F., You-Sheng L. et al. $T_c$ and the elastocaloric effect of $Sr_2RuO_4$ under <110> uniaxial stress: No indications of transition splitting. 2024. Phys. Rev. B 110, 064514 DOI: https://doi.org/10.1103/PhysRevB.110.064514
[59] Hicks, CW., Jerzembeck, F., Noad, H., Barber, M. & Mackenzie, A. Probing Quantum Materials with Uniaxial Stress. 2025. Vol. 16, pp. 417–442, https://doi.org/10.1146/annurev-conmatphys-040521-041041
[60] Scalapino. D. 1995, The case for $d_{x2-y2}$ pairing in the cuprate superconductors Physics Reports. 1995. 250 (6): 329-365, https://doi.org/10.1016/0370-1573(94)00086-I
[61] Bednorz, J. & Müller. K. Possible high Tc superconductivity in the BaLaCuO System. 1986. Z. Physik B - Condensed Matter 64, 189–193. https://doi.org/10.1007/BF01303701
[62] Contreras, P. & Moreno, J. Nonlinear minimization calculation of the renormalized frequency in dirty d-wave superconductors. 2019. Can. J. Pure Appl. Sci. 13(2):4807-4812. ISSN: 1920-3853.
[63] Contreras, P. and Osorio, D. Scattering due to non-magnetic disorder in 2D anisotropic d-wave high Tc superconductors. 2021. Engineering Physics. 5(1):1-7 https://doi.org/10.11648/j.ep.20210501.11